\documentclass[sigplan,10pt]{acmart}
\acmConference[Technical Report]{Technical Report (PPoPP Template)}{September 2026}{}
\acmYear{2026}
\acmISBN{}
\acmDOI{}
\startPage{1}
\setcopyright{none}
\usepackage{booktabs}
\usepackage{subcaption}
\usepackage{tabularx}
\usepackage{listings}
\usepackage{xurl}
\usepackage{tikz}
\usetikzlibrary{arrows.meta,positioning,fit,shapes.geometric}
\usepackage[nameinlink,noabbrev]{cleveref}
\hypersetup{
  pdfauthor={Siming HUANG},
  pdfsubject={Correctness recovery and inference performance engineering for DeepSeek-V4-Flash on AMD gfx90a},
  pdfkeywords={LLM inference, mixture of experts, AMD MI250, tensor parallelism, quantization, correctness}
}

\definecolor{mfblue}{HTML}{2F5D8C}
\definecolor{mfteal}{HTML}{2A8C82}
\definecolor{mforange}{HTML}{D17A3A}
\definecolor{mfred}{HTML}{B44A4A}
\definecolor{ink}{HTML}{25313C}
\definecolor{codebg}{HTML}{F5F7FA}
\lstdefinestyle{reportcode}{
  basicstyle=\ttfamily\small,
  backgroundcolor=\color{codebg},
  frame=single,
  rulecolor=\color{black!15},
  breaklines=true,
  columns=fullflexible,
  keepspaces=true,
  showstringspaces=false,
  upquote=true
}
\newcolumntype{Y}{>{\raggedright\arraybackslash}X}

\newcommand{\risk}[1]{\textbf{#1}}
\newcommand{\commit}[1]{\texttt{#1}}
\newcommand{\DecodeOne}{88.63}
\newcommand{\DecodeThirtyTwo}{1041.78}
\newcommand{\DecodeSixtyFour}{1327.10}
\newcommand{\PrefillMin}{7035.83}
\newcommand{\PrefillMax}{10431.43}
\newcommand{\PrefillEight}{10413.87}
\newcommand{\PrefillSixteen}{10263.02}
\newcommand{\PrefillThirtyTwo}{9966.84}
\newcommand{\PeakMemoryMB}{54743}
\newcommand{\PeakMemoryPercent}{83.55}

\begin{document}

\title[DeepSeek-V4-Flash on AMD gfx90a]{DeepSeek-V4-Flash on AMD gfx90a}
\subtitle{Correctness Recovery and Inference Performance Engineering}
\author{Siming HUANG}
\affiliation{
  \institution{HKUST(GZ)}
  \city{Guangzhou}
  \country{China}
}

\begin{abstract}
We report the enablement, correctness recovery, and performance engineering of DeepSeek-V4-Flash inference on AMD Instinct MI250 (gfx90a/CDNA2) using SGLang. Original mixed FP4/FP8 checkpoint weights are retained; execution uses shape-specific HIP, AIter, Composable Kernel, and Triton paths. The study extends an earlier four-GCD investigation to a single eight-GCD tensor-parallel instance with a 1,048,576-token logical KV pool. A fresh-process native autoregressive campaign on public-source code requests covers concurrency 1 through 64, with three measured rounds per group. Single-request resident decode reaches \DecodeOne{} output tokens/s, while C32 and C64 reach \DecodeThirtyTwo{} and \DecodeSixtyFour{}. Separate C16 prefill measurements reach \PrefillEight{}, \PrefillSixteen{}, and \PrefillThirtyTwo{} input tokens/s at approximately 8K, 16K, and 32K input length. The accepted prefill configuration combines reduced indexer replication, mHC reuse, paired-GCD attention, and fixed-order expert reduction. We distinguish current measurements from historical ablations and speculative throughput, and resident decode from whole-request HTTP latency. Component equality and bounded semantic checks do not establish universal numerical equivalence; dynamic batching, cold-shape compilation, and untested million-token occupancy remain explicit limitations.
\end{abstract}

\keywords{LLM inference, mixture of experts, AMD MI250, tensor parallelism,
  quantization, correctness, speculative decoding}
\maketitle
\hypersetup{
  pdfauthor={Siming HUANG},
  pdfsubject={Correctness recovery and inference performance engineering for DeepSeek-V4-Flash on AMD gfx90a},
  pdfkeywords={LLM inference, mixture of experts, AMD MI250, tensor parallelism, quantization, correctness}
}

\section{Executive summary}

This report updates the evidence through September 18, 2026. The current
campaign uses execution HEAD \commit{147ce11b25}, recorded working-tree hashes,
and the explicitly enabled accepted prefill configuration. The September 14
matrix archive \commit{3d73a75923} and pilot \commit{fb2e39fca9} remain
historical evidence, not freshly repeated ablations.

\begin{table*}[tp]
  \centering
  \caption{Evidence-backed state as of September 18. C denotes client concurrency; resident decode excludes admission and drain.}
  \label{tab:headline}
  \begin{tabularx}{\textwidth}{@{}p{0.29\textwidth}Y@{}}
    \toprule
    Item & Result \\
    \midrule
    Model & DeepSeek-V4-Flash; 284B total parameters, approximately 13B active parameters, FP4 routed experts, and mostly FP8 non-expert weights \\
    Host platform & Supermicro A+ Server AS-4124GQ-TNMI; dual AMD EPYC 7763 processors with SMT enabled; 1 TiB DDR4-2933 from 32 $\times$ 32 GiB 2Rx8 RDIMMs \\
    Accelerators & Eight MI250 GCDs (gfx90a), approximately 64 GiB HBM per GCD; four dual-GCD accelerators, not eight MI250 boards \\
    Formal native decode & C1 \DecodeOne{}; C32 \DecodeThirtyTwo{}; C64 \DecodeSixtyFour{} resident output tok/s; full seven-tier table in \cref{tab:tp8-matrix} \\
    Formal prefill & 8K public-source inputs: \PrefillMin{}--\PrefillMax{} input tok/s across C1--64; admission limited to 16 requests \\
    Long-input prefill & C16: \PrefillSixteen{} at 16K and \PrefillThirtyTwo{} at 32K input tokens/request \\
    KV pool & 1,048,576 logical tokens allocated; no filled-1M-context performance or quality certification \\
    Runtime topology & One TP8/EP1 instance, no routed-expert A2A; native decode graph tiers 1/2/4/8/16/32/64 \\
    Service state & Benchmark services stopped after testing; no live endpoint is implied by this report \\
    \bottomrule
  \end{tabularx}
\end{table*}

The report applies three evidence rules:

\begin{enumerate}
  \item Native AR and full-target DSpark are distinct from historical approximate verification. Their rates are not interchangeable.
  \item Original checkpoint precision, component equality, repeated output hashes, and end-to-end answer quality are distinct claims. No one check substitutes for the others.
  \item Formal three-round matrices and small ABBA screens remain separate. Minima/maxima describe observed runs, not confidence intervals; percentages from different experiments are not added together.
\end{enumerate}

\section{Related systems}

This work builds on serving systems that reduce scheduling and memory-management overhead for autoregressive language models. SGLang combines a structured runtime with RadixAttention and continuous batching, while vLLM introduced PagedAttention to manage KV-cache fragmentation and sharing efficiently~\cite{sglang,vllm}. Our contribution is narrower and hardware-specific: we retain SGLang's serving abstractions while adapting the DeepSeek-V4 execution path to CDNA2 and validating optimizations against fixed-token numerical oracles.

DeepSeek-V3 established the MoE, routing, multi-token prediction, and low-precision training lineage; DeepSeek-V4 adds compressed sparse/heavily compressed attention and mHC~\cite{dsv3,dsv4report}. MegaBlocks demonstrates block-sparse MoE training~\cite{megablocks}; our serving workload instead spans tiny per-expert decode matrices and large prefill chunks. These regimes require different execution geometries rather than a universal kernel. DSpark supplies a separate confidence-scheduled speculative design~\cite{dspark}; we report its historical full-target results separately from the current native-AR matrix.

On AMD GPUs, Composable Kernel and AIter provide tuned HIP, CK, and CKTile building blocks for GEMM, quantization, attention, and MoE~\cite{ck,aiter}. We use these libraries where their supported shapes are effective, but introduce gfx90a-specific paths where generic kernels do not match FP4 storage, wave64 execution, or the small-$M$ decode regime. The resulting system is therefore an integration and measurement study rather than a replacement for either the serving runtime or the kernel libraries.

\section{System and model context}

\subsection{Hardware and software stack}

The experimental host is a Supermicro A+ Server AS-4124GQ-TNMI with two AMD EPYC 7763 processors and SMT enabled. Host memory comprises 32 32-GiB DDR4-2933 2Rx8 RDIMMs, totaling 1 TiB. Historical TP4 runs used four GCDs; the new TP8 matrix uses all eight GCDs as one model instance, not two independent TP4 replicas. The accelerators are MI250, not MI250X.

The recorded software environment is:
\begin{itemize}
  \item Python 3.12.13;
  \item PyTorch \path{2.12.0a0+git78258b9};
  \item Triton \path{3.7.1+git0263a6a6.rocm7.14.0};
  \item Transformers 5.12.1 and Tokenizers 0.22.2.
\end{itemize}
The installed HIP compiler reports 7.15.26333 and AMD clang 23; this is distinct from the HIP 7.14 build provenance recorded in earlier runs and does not identify every cached binary. SGLang, AIter, and CK include local source changes. The environment ledger lists dependency commits and dirty paths; a clean SGLang commit alone is not a complete binary reproducibility guarantee.

The original checkpoint comprises 48 indexed safetensors shards. Recorded SHA256 values cover configuration, tokenizer, generation settings, and the shard index, not every weight byte. This study does not include V4.1 or Engram offload results.

CDNA2 provides wave64 execution, 64 KiB LDS per CU, and INT8/BF16/FP16 matrix or dot instructions, but no native FP4 matrix core comparable to newer accelerator generations. DeepSeek-V4-Flash stores routed-expert weights in FP4. Consequently, the critical gfx90a cost is not raw HBM bandwidth alone: it is the combination of small-M utilization, online FP4 unpacking, scale handling, activation quantization, expert sorting, and layer-level synchronization. The AMD MI200 ISA manual guided the custom HIP kernels in this work~\cite{amdisa}.

\subsection{Execution structure}

DeepSeek-V4-Flash has 43 decoder layers, 256 routed experts, top-6 routing per token, shared experts, mHC, and compressed sparse attention. The official model card reports 284B total parameters, 13B active parameters, a one-million-token context, and mixed FP4 expert / FP8 non-expert precision~\cite{dsv4card}.

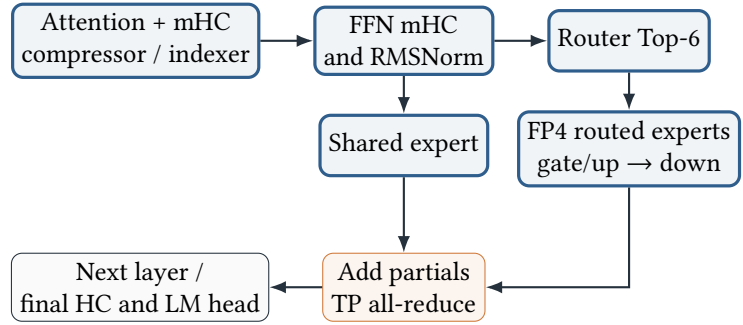
\begin{figure*}[tp]
  \centering
  \begin{tikzpicture}[
    node distance=5mm and 7mm,
    box/.style={draw=ink,rounded corners,align=center,minimum height=8mm,fill=black!2},
    hot/.style={draw=mfblue,very thick,rounded corners,align=center,minimum height=8mm,fill=mfblue!7},
    comm/.style={draw=mforange,rounded corners,align=center,minimum height=8mm,fill=mforange!8},
    arrow/.style={-{Latex[length=2.2mm]},thick,color=ink}
  ]
    \node[hot] (attn) {Attention + mHC\\compressor / indexer};
    \node[hot,right=of attn] (ffn) {FFN mHC\\and RMSNorm};
    \node[hot,right=of ffn] (router) {Router Top-6};
    \node[hot,below=of router] (routed) {FP4 routed experts\\gate/up $\rightarrow$ down};
    \node[hot,below=of ffn] (shared) {Shared expert};
    \node[comm,below=10mm of shared] (reduce) {Add partials\\TP all-reduce};
    \node[box,left=of reduce] (head) {Next layer /\\final HC and LM head};
    \draw[arrow] (attn) -- (ffn);
    \draw[arrow] (ffn) -- (router);
    \draw[arrow] (router) -- (routed);
    \draw[arrow] (ffn) -- (shared);
    \draw[arrow] (routed.south) |- (reduce.east);
    \draw[arrow] (shared) -- (reduce);
    \draw[arrow] (reduce) -- (head);
  \end{tikzpicture}
  \caption{Schematic native TP-only layer boundaries. Shared and routed experts contribute to a join. TP4/EP1 and TP8/EP1 avoid expert A2A but retain TP collectives; the historical EP variants instead require dispatch/combine.}
  \label{fig:architecture}
  \Description{Native TP-only dataflow: attention feeds FFN normalization, which branches to shared and routed experts; their partials join before a TP all-reduce.}
\end{figure*}

\section{gfx90a enablement}

The initial ROCm implementation targeted gfx942 and gfx950. Enabling gfx90a required the following changes:

\begin{itemize}
  \item add the gfx90a build target and select \path{HIP_FP8_TYPE_FNUZ};
  \item repair ROCm 7.14 include, library, and rpath discovery while avoiding the conda host-side \texttt{hipcc};
  \item build AOT HIP kernels and enable AIter FP4 CK MoE execution on gfx90a;
  \item handle Mori topology and XGMI initialization on a node without an RDMA NIC;
  \item validate graph capture and replay on gfx90a, separating kernel selection failures from hardware faults;
  \item establish both TP4/EP4 with Mori and TP4/EP1 without A2A as executable reference paths.
\end{itemize}

The initial bring-up checkpoint was \commit{505b3373794a}. It predates the later correctness and performance work and must not be treated as the final state.

\section{Correctness recovery}

\subsection{Admission, sampling, and chat encoding}

An early request stalled at SWA admission, not in Mori or RCCL. A 4096-token pool, SWA ratio 0.1, and page size 256 together fell below the admission floor. A separate AIter sampler issue on gfx90a could return an invalid token ID on special logits rows; the replacement uses \path{torch.argmax} for this case.

DeepSeek-V4 does not ship a Jinja chat template. It requires the model-specific encoding and output parser, which is also stated in the official model card~\cite{dsv4card}.

\subsection{W2 permutation: invalidating the early 60 tok/s result}

\risk{The early approximately 60 tok/s AIter FP4 route was not correctness-valid.} The legacy CK FP4-by-FP4 stage-1 path produced zeros on gfx90a. After switching to CKTile BF16-by-FP4, stage 1 reached approximately 0.999996 cosine similarity to a direct oracle, but stage-2 W2 remained near 0.24.

Output-column fingerprinting identified the following 16-row block mapping within each 128-wide N tile:

\begin{equation}
  \text{fast}\rightarrow\text{reference}=[0,2,4,6,1,3,5,7].
\end{equation}

Applying the inverse permutation

\begin{equation}
  [0,4,1,5,2,6,3,7]
\end{equation}

to raw W2 weights and scales during loading restored the logical output order for top-k=1 and top-k=6. This is a one-time load transformation with no decode hot-path cost.

The fixed France oracle uses the following input IDs:

\begin{lstlisting}
[0,128803,3085,344,270,6102,
 294,8760,2755,128804,128822]
\end{lstlisting}

Its expected completion is:

\begin{lstlisting}
The capital of France is **Paris**.
token ids: [671,6102,294,8760,344,2619,51119,42499,1]
hash: 6f41fe2f01d52507
\end{lstlisting}

The short sentinel was stable in the documented eager, graph, and independent-start checks. This bounded test cannot certify a long context, a different batch shape, or the entire numerical pipeline. Greedy completions can diverge from the first token under different prefill execution; fixed-input logits and layer comparisons are needed to locate such differences.

\subsection{Later numerical and integration repairs}

The later investigation also identified an independent E8M0 scale-layout issue in large-prefill BF16-CK execution: raw packed weights do not imply raw logical scales. Runtime AIter scale shuffling must be inverted or consumed with the matching address map. Constant-scale synthetic fixtures can hide this error. The latest scoped large-prefill path uses unique-slot stores and fixed Top-6 FP32 reduction; other fallback paths may still use atomics. Neither scale-layout repair nor a fixed final reduction proves whole-model batch invariance.

A subsequent TP mapping repair passes the rank-local attention sink rather
than the full checkpoint vector to the H8 kernel. Previously, nonzero ranks
could read rank 0's head sinks. Equality to that old kernel was therefore not
a valid model reference. The paired-GCD H16 path used in the latest campaign
is based on the corrected mapping. Further mHC repairs retain original FP32
coefficient weights in the scoped first/unfused prefill boundaries. The first
boundary's static 24-value projection can be tabulated by vocabulary ID,
without replacing context-dependent residuals or later-layer coefficients.

Deterministic index selection fixes membership ties using logical IDs and presents selected positions in a fixed order before rank-local physical mapping. The prefill trivial-row shortcut retains real sequence lengths and all compressor/cache updates; only scores irrelevant to Top-K are skipped. Subsequent native-decode empty-tile elimination is a separate optimization, described in \cref{sec:current-native}.

Fresh starts after the upstream/V4.1 integration exposed missing C4 lifecycle callbacks and incorrect compression-ratio discovery for unified V4 pools. Their repair restored the requested logical 1M pool and native graph construction. CPU tests also cover SWA-only draft metadata, but the latest user scope excluded DSpark: the completed native matrix is not a fresh DSpark GPU validation.

\section{Historical TP4 decode optimization}

\subsection{Parallel decomposition}

After the W2 repair, the corrected TP4/EP4+Mori BS1 baseline was only 14.2--15.0 tok/s. The TP4/EP1 no-A2A CKTile reference reached approximately 20.14 tok/s. This comparison motivated investigation of dispatch/combine, rank imbalance, and progress overhead; changing the expert decomposition also changes compute shapes, so it does not isolate communication cost.

The historical TP4/EP1 decode route combines:

\begin{itemize}
  \item packed 4-bit FP4 weights without an offline expansion;
  \item online BF16 activation quantization to INT8 per 32 elements;
  \item E2M1 nibble mapping to $0,\pm1,\pm2,\pm3,\pm4,\pm6,\pm8,\pm12$;
  \item CDNA2 \texttt{v\_dot4\_i32\_i8} / mixed-dot accumulation followed by activation and E8M0 scaling;
  \item four 16-lane subgroups for top-6 slots in the down projection;
  \item AIter peer-read custom all-reduce in place of the high fixed-latency RCCL path;
  \item TP-only mHC launch geometry that no longer reserves CUs for Mori progress.
\end{itemize}

\begin{figure*}[tp]
  \centering
  \includegraphics[width=0.96\textwidth]{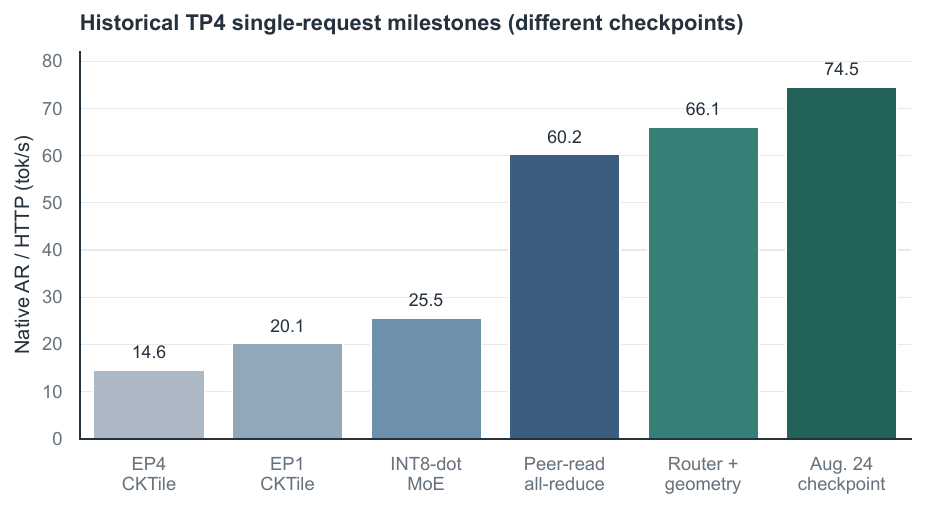}
  \caption{Historical TP4 short-probe milestones retained from the earlier report. Different checkpoints and controls are not a single controlled speedup series. The invalid early FP4 route is excluded; the later 60.20 point follows a separate repair.}
  \label{fig:decode-progress}
  \Description{Historical TP4 single-request native HTTP rates rise from 14.6 to 74.5 tokens per second across six different checkpoints; these are not one controlled ablation.}
\end{figure*}

The ledger associates the largest jump in \cref{fig:decode-progress} with peer-read all-reduce, approximately 25.54 to 60.20 tok/s. The August short-probe milestone is about 74.5 tok/s. A September 6 isolated-tree reproduction obtained a seven-request HTTP median of 74.018 and maximum of 74.655 tok/s, with the historical hash on all seven requests. This was \emph{TP4 native AR}, not TP8 or DSpark. The inherited rebased tree ran about 53 tok/s; its different tree contents exposed a lost gfx90a MHC selector. Historical author dates and copied experiment notes did not establish performance on the rebased code. The current TP8 C1 result uses a different public-code protocol and must not be divided by 74.5 to claim TP scaling efficiency.

\section{Prefill optimization and historical checkpoints}

\subsection{From per-assignment scalar work to CDNA2 MFMA}

The original raw-FP4 direct kernel sent M=256 prefill through a per-assignment FP16 dot path, with approximately 12.9 s TTFT for a 1028-token prompt. The optimization sequence introduced:

\begin{enumerate}
  \item packed-FP4 unpack reuse across rows, first group8 and then group32 for M$\geq$1024;
  \item \texttt{v\_perm\_b32} lookup tables instead of branch-heavy nibble decoding;
  \item MI250 block-FP8 tuned configurations for M512 and M1024 dense projections;
  \item CDNA2 INT8 MFMA routed gate and down kernels;
  \item sparse paged-prefill attention reduction from eight waves to one wave;
  \item chunked-prefill growth from 512/1024 to 2048;
  \item M2048-specific MFMA grids, native HIP INT8 quantization, and scale/metadata broadcast;
  \item an expert sorter block increase from 32 to 64 rows, allowing an expert with approximately 48 assignments to scan packed weights once.
\end{enumerate}

\begin{figure*}[tp]
  \centering
  \includegraphics[width=0.96\textwidth]{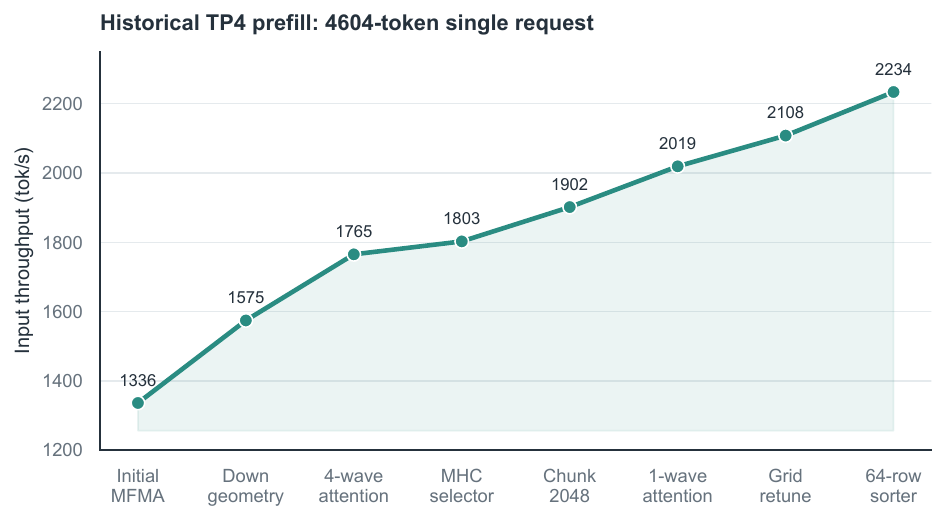}
  \caption{Historical TP4 sequence for a 4604-token prompt. Rates derive from HTTP TTFT; these points are not the latest 8K-input TP8 matrix.}
  \label{fig:prefill-progress}
  \Description{Historical 4604-token TP4 prefill throughput rises from 1336 to 2234 input tokens per second across eight chronological configurations.}
\end{figure*}

\begin{figure*}[tp]
  \centering
  \begin{minipage}[t]{0.49\textwidth}
    \centering
    \includegraphics[width=\linewidth]{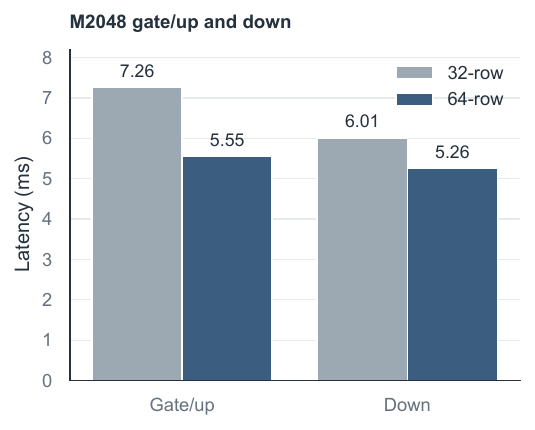}
  \end{minipage}\hfill
  \begin{minipage}[t]{0.49\textwidth}
    \centering
    \includegraphics[width=\linewidth]{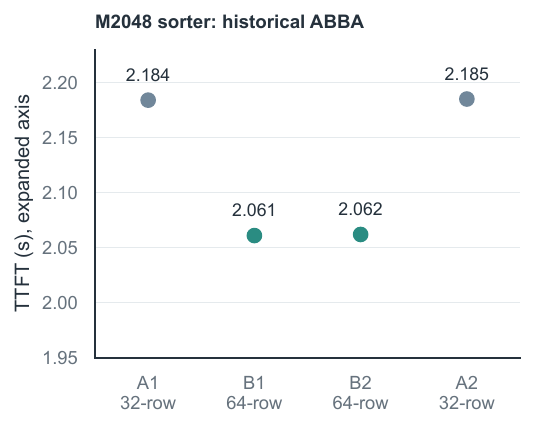}
  \end{minipage}
  \caption{Left: M2048 routed-MoE kernel latency. Right: strict endpoint-level ABBA validation of the 64-row sorter.}
  \label{fig:micro}
  \Description{The 64-row sorter reduces gate and down microbenchmark latency. The separate endpoint ABBA records 2.184, 2.061, 2.062 and 2.185 seconds.}
\end{figure*}

The historical M2048 microbenchmark compares the 32-row and 64-row sorter variants in \cref{fig:micro}. Gate/up improves from 7.26 to 5.55 ms (23.6\%), down improves from 6.01 to 5.26 ms (12.5\%), and the combined kernel time improves by approximately 18.5\%. The recorded component fixtures reported identical outputs; this is not a full-model equality claim.

The historical endpoint ABBA sequence was A1=2.184 s, B1=2.061 s, B2=2.062 s, and A2=2.185 s. A later chunk-only TP4 test used B1=1.861 s, A=1.978 s, and B2=1.852 s (six steady samples per arm). Chunk 2304 removes the small tail traversal for 4604 tokens; it is not a universal optimum over prompt lengths. The first new-shape request still took over 20 s, whereas the warm measurements were near 1.85 s.

\subsection{Large-M CK and the meaning of the earlier 6.42k result}

The subsequent BF16-CK prefill route materializes the current expert shard in BF16, runs CK stages, and casts the accumulated output back to BF16. It does not alter the checkpoint files, but changes execution arithmetic relative to raw-FP4/INT8-dot kernels. In the September 7 TP8 migration, 32 distinct requests totaling 73,724 input tokens, admission 16, and two approximately M36864 forwards reached a warm median of 6420.39 input tok/s with a 131,072-token pool. This is an aggregate prefill wave, not single-request decode. The September 14 matrix used approximately 8K input per request and a 1M pool, reaching 4.68--5.27k. Its different length, pool, grouping, and timing prevented a direct regression claim. The September 18 matrix below replaces that historical matrix as the current configuration's measurement.

\section{Current TP8 native-AR results}
\label{sec:current-native}

The primary result is a fresh-process remeasurement on September 18 at
execution HEAD \commit{147ce11b25}, with working-tree and launch hashes retained.
The same process runs the cells serially, with explicit per-shape warmups and
three measured rounds per cell. All seven native graph tiers have matching
active input and executed rows during full residency. The model remains a
single TP8/EP1 instance. No DSpark, MTP, or approximate routed-expert
verification contributes to the following table. Earlier TP4, speculative,
and individual optimization ablations remain historical records.

\begin{table*}[tp]
  \centering
  \caption{Formal native P/D matrix: P and resident D are three-round medians;
  HTTP aggregates complete measured waves. Prefill divides aggregate input by
  wave time to the last first token. Resident and HTTP decode are distinct metrics.}
  \label{tab:tp8-matrix}
\begin{tabular}{@{}rrrr@{}}
\toprule
C & Prefill input tok/s & Resident output tok/s & HTTP output tok/s \\
\midrule
1 & 7035.83 & 88.63 & 86.76 \\
2 & 9176.62 & 107.41 & 102.89 \\
4 & 10278.75 & 187.13 & 152.72 \\
8 & 10362.67 & 333.38 & 250.49 \\
16 & 10413.87 & 598.11 & 458.78 \\
32 & 10431.43 & 1041.78 & 689.74 \\
64 & 10381.49 & 1327.10 & 829.65 \\
\bottomrule
\end{tabular}

\end{table*}

\begin{figure*}[tp]
  \centering
  \includegraphics[width=\textwidth]{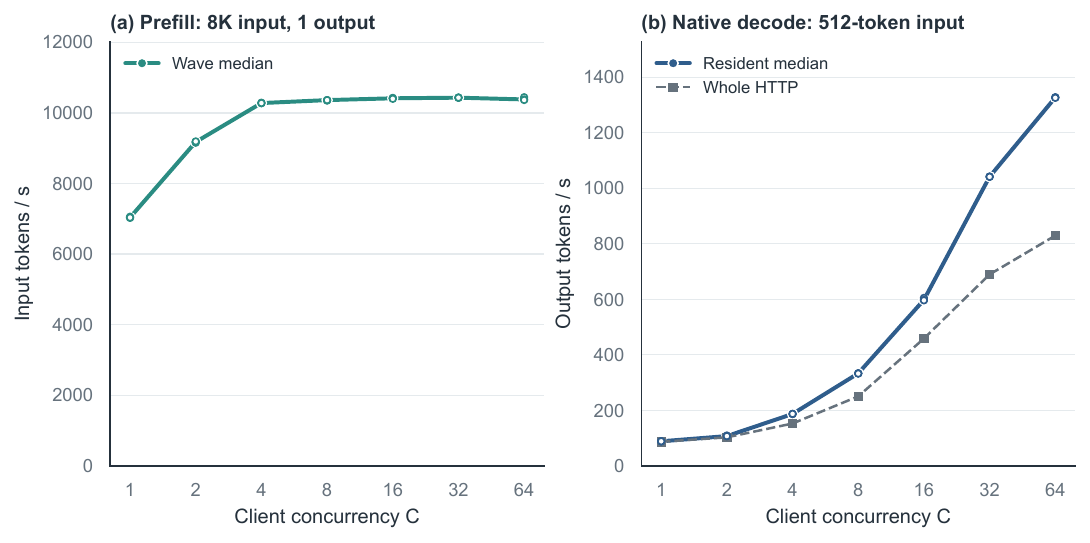}
  \caption{Latest TP8 matrix. Colored points/lines show three-round medians;
  thin vertical ranges and small points are actual rounds, not confidence
  intervals. Gray decode points are complete-wave HTTP rates, not a slower
  kernel configuration. Concurrency uses a base-2 axis.}
  \label{fig:tp8-pd}
  \Description{Current TP8 concurrency 1 through 64: prefill rises from its single-request rate to a plateau above ten thousand input tokens per second; native resident decode and whole-request HTTP rates are shown separately.}
\end{figure*}

\paragraph{Protocol boundaries.}
Prefill uses 8191--8192-token public-source prompts, fresh cache salts, and one
output token. At C64 the wave contains about 524k input tokens, admitted at
most 16 requests at a time with a 32,768-token chunk budget. C64 therefore does
not mean one M64 prefill kernel or 64 simultaneous large prefill requests.
Decode uses 511--512-token inputs, greedy sampling, natural EOS, and at most
2048 output tokens. Each formal round accumulates at least 30 s of common
resident windows; explicit warmups are excluded. The 64-case corpus extends
the earlier 32-case public-source corpus without changing those first 32 cases.
Duration-dependent wave counts can expose different case mixtures, which are
preserved in raw records.

\paragraph{One consistent launch configuration.}
The current run enables the accepted C1-only wo\_a GEMV for the entire matrix,
including larger-concurrency drain periods. Unlike the September 14 table,
the new C1 cell is not spliced in from a separate supplement. The default-off
native C32 down-consumer pilot is not enabled. The rejected fixed-Lt
owner-only query producer is explicitly disabled. Prefill uses the scoped
accepted configuration described below, not an assertion that every launcher
default automatically selects the same path.

\paragraph{Interpretation.}
The 8K input rates range from \PrefillMin{} to \PrefillMax{} tok/s,
with C16 at \PrefillEight{}. The native resident decode endpoints are
\DecodeOne{} at C1 and \DecodeSixtyFour{} at C64. The HTTP/resident gap includes
admission, prefill, uneven answer lengths, and drain; it is not a measured
host-only overhead term. All measured rounds, including outliers, are retained.
The older matrix is preserved in the frozen data, but old/new ratios are not
controlled cumulative speedups: the chunk budget, correctness repairs, and
active numerical paths changed. No matched current TP4 seven-tier matrix is
available, so a TP4-to-TP8 scaling factor is not reported.

\subsection{Accepted large-prefill configuration and longer inputs}

The current configuration combines several independently developed changes:
causal empty-tile elimination; query-group reuse and runtime-$M$ indexer
kernels; row-owner score/Top-K work with logical-index exchange; exact paired
FP32 mHC pre-mix ownership and HIP post-update; and temporary H8-to-H16
attention activation redistribution within paired GCDs. Full indexer query
production and compressor/cache updates remain intact. Wide-owner paths
cover the measured 16K and 32K inputs without truncating their KV selections.

For routed MoE, the accepted large-prefill path retains original checkpoint
storage, uses layer-local BF16 expansion, and combines route production with
unique-slot CK stage-2 stores and a vectorized fixed-order Top-6 FP32 reducer.
This eliminates the final cross-expert atomic order in the admitted path;
it does not establish batch-invariant projections or universal determinism.
Original FP32 mHC coefficients are retained at the corrected first/unfused
boundaries. A small static first-layer vocabulary table replaces only its
24-value raw pre-mix projection, not subsequent context-dependent operations.

\begin{table}[tp]
  \centering
  \caption{Current C16 length sweep, three-round medians. Original weights,
  1M allocated logical KV, 32K chunks, zero prefix-cache hits.}
  \label{tab:current-lengths}
  \begin{tabular}{@{}rr@{}}
    \toprule
    Input tokens/request & Aggregate input tok/s \\
    \midrule
    Approximately 8K & \PrefillEight{} \\
    Approximately 16K & \PrefillSixteen{} \\
    Approximately 32K & \PrefillThirtyTwo{} \\
    \bottomrule
  \end{tabular}
\end{table}

These are separate actual-length workloads, not a same-shape regression or
a filled-million-token-context test. The experiment records resolved settings,
eight-rank backend witnesses, source hashes, input echoes, zero cache-hit counts,
and raw timing records. France and ID/text checks provide
bounded validation, not a task-accuracy benchmark. Historical ablation gains
are not added together to explain the combined configuration.

The current three-round C16 prefill tests also expose a repeatability limit:
two of the sixteen 16K requests change their first token across rounds;
the corresponding 8K and 32K C16 first-token sets repeat exactly. These are
one-token observations, not full-answer equality tests. The differing 16K
responses are retained in the timing data. Fixed CK reduction and local
component equality must not be presented as whole-model determinism.

\subsection{Historical ablation: empty C4 tiles}

With a 1M logical pool, the C4 logits launch could cover 262,144 key positions
even when only a short prefix was active. Beyond the raw-token threshold near
2048, many launched tiles had no valid keys but still performed unnecessary
query loading and dot work. The exact native-decode guard skips this empty
work while retaining defined zero stores, unchanged active computation,
sequence lengths, Top-K semantics, and physical index mapping.

Seven component shapes, 100 mutations per shape, and 1000 graph replays per
shape checked score bits and logical/physical indices. Fully populated control
cases did not exhibit the empty-work benefit. At M32, the diagnostic component
dropped from approximately 6934 to 837 microseconds; this is not a whole-model
speed prediction.

A separate matched 8K-input C32 service ABBA measured
169.657 / 611.103 / 610.853 / 169.843 resident output tok/s, a 3.599-fold
candidate/control ratio. Complete HTTP rates were approximately 83.18 / 128.85 /
128.88 / 83.26 tok/s. The different ratios illustrate why the measurement
boundary must be explicit. This 8K test is not the 512-input formal decode
table, and its multiplier must not be applied to that table. This fix is
native-decode scoped; the earlier prefill trivial-row skip is a different path.

\subsection{Historical ablation: C1 projection specialization}

The restored wave64 wo\_a GEMV measured 30.75$\rightarrow$6.88 microseconds in
its component test. All 100 tested mutations were finite and repeatable, but
only 70 were bit-equal to the former einsum path; the maximum relative L2
error against a float32 reference was about 0.00179 on the test inputs.
It retains checkpoint weights while changing floating reduction.

The service ABBA was 79.558 / 87.583 / 87.504 / 79.151 resident output tok/s:
+10.32\% by mean candidate/control arms. Candidate repeats had identical
completion sequences in that test, and France passed. The separate formal C1
median is 87.599. These controls are more informative than an unpaired
comparison with the August 74.5 HTTP result.

\begin{figure*}[tp]
  \centering
  \includegraphics[width=\textwidth]{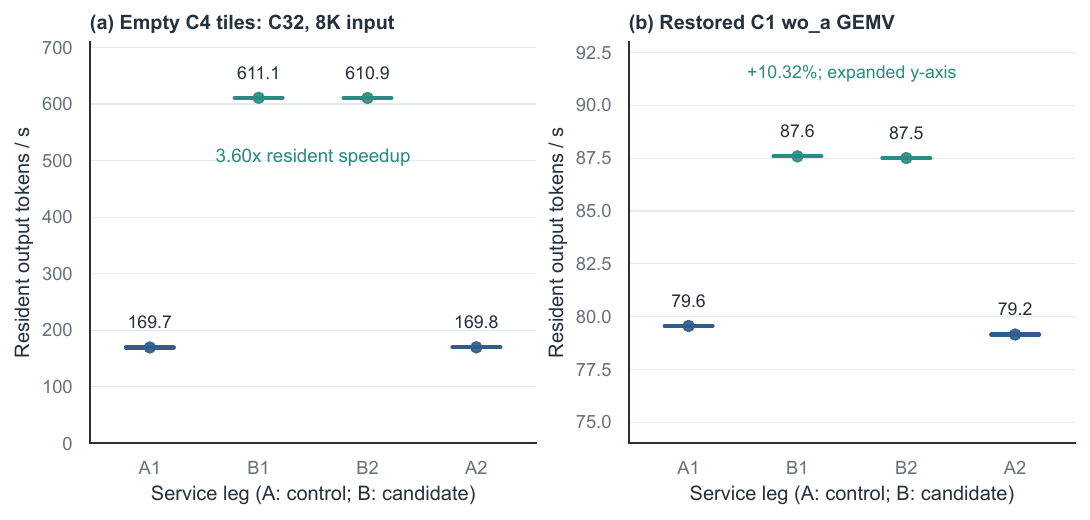}
  \caption{Separate native ABBA experiments: empty C4 tiles at C32 with 8K
  inputs, and the C1 projection specialization with 512-token code inputs.
  The right axis is expanded; the experiments must not be multiplied together.}
  \label{fig:native-abba}
  \Description{Separate ABBA plots show a 3.60-fold resident gain for empty C4 tile elimination at C32 with 8K inputs, and a 10.32 percent gain for the C1 GEMV specialization.}
\end{figure*}

\subsection{Historical down-consumer pilot: a small C32 gain}

The follow-up at \commit{fb2e39fca9} replaces the M32 second quantization/down
chain with a CTA16 consumer. Each CTA reads the already-rounded BF16
intermediate, forms group32 INT8 values/scales in LDS, computes its output
stripe, and retains the existing FP32 top-k partial and fixed reduction.
The current gate, sorter, original weights, attention, and TP collectives are
unchanged. The selector is limited to native TP8/EP1, M32/I256 and A4/R2
geometry; it remains default-off.

On one GCD, synthetic diverse/skewed inputs reduced quant+down+reduction from
130.542 to 115.907 microseconds and 81.615 to 70.791 microseconds, respectively.
Each distribution passed 100 mutations of inputs, routes, weights and scales
and 1000 checked graph replays. Comparisons used numeric element equality,
not a signed-zero-bit check. This is neither the full routed stage nor a
captured real-layer oracle.

The service screen uses two natural-EOS waves per leg with the same 32 code
requests, rather than the duration-driven formal 64-case protocol. C32 leg
medians are 1047.763 / 1060.206 / 1061.203 / 1041.459 resident tok/s. Mean
control 1044.611 versus candidate 1060.705 gives +1.541\%; the return control
changes $-0.602\%$, and candidate legs differ 0.094\%. Every candidate wave
exceeds every control wave. Whole HTTP rates vary with answer length and do
not establish the same percentage gain.

The conditional C1 test is an isolation check, not an M1 port: the existing
direct M1 down kernel already quantizes in LDS, with a different rounding
contract. Its control/candidate means are 89.722/90.024 tok/s; the 0.34\%
variation is not attributed to an unselected kernel. All eight 1453-token C1
completions match across arms. This fixed-case screen does not supersede the
formal multi-case C1 value. For C32, control outputs themselves vary across
waves; neither whole-model bitwise parity nor code-answer factual correctness
is established. All 264 measured pilot responses passed independent ID/text
integrity checks, and all service France sentinels passed.

\begin{figure*}[tp]
  \centering
  \includegraphics[width=\textwidth]{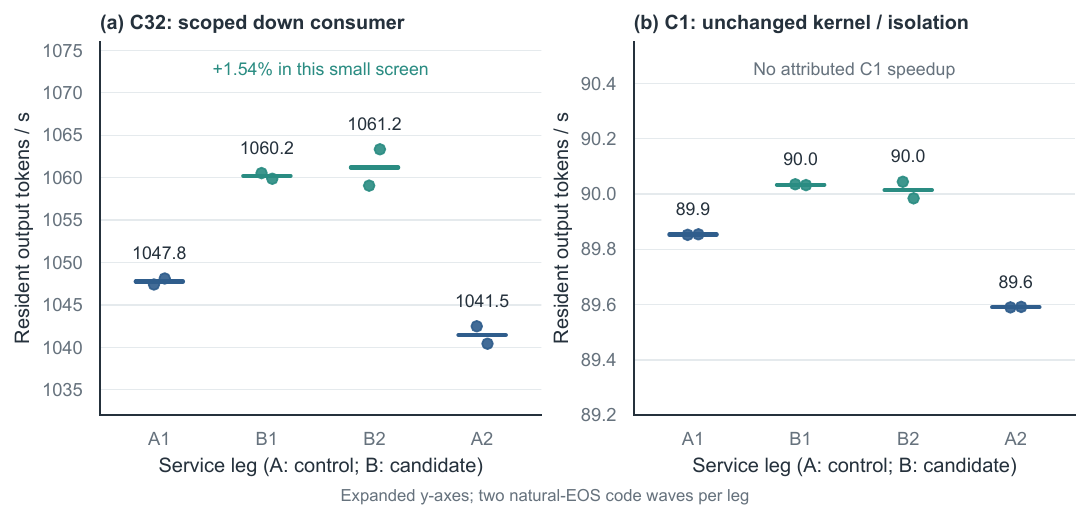}
  \caption{Default-off down-consumer pilot, expanded y-axes. Dots are the two
  measured waves per leg; short bars mark their median. C1 never executes
  the candidate and is a negative-control scope check.}
  \label{fig:consumer-abba}
  \Description{Two waves per ABBA leg show a 1.54 percent resident C32 consumer gain; C1 is an unchanged-kernel isolation check with no attributed gain. Both axes are expanded.}
\end{figure*}

\section{Historical speculative decoding: separate evidence}
\label{sec:spec-history}

The latest native-AR matrix deliberately excludes DSpark. We retain the
September 11 full-target result as historical evidence, not a claim that the
latest merged DSpark service has been revalidated. This distinction also
prevents an earlier approximate-target speed from becoming a recovery target.

\subsection{Why the historical TP4 1.5k result is not a strict baseline}

In the old anchor-only/compact routed-MoE path, C32 with three draft positions
produced M128 target rows, but only 32 anchor rows executed the full routed
expert branch. Non-anchor rows omitted routed experts. Keeping stored weights
unchanged did not preserve the full target function at those positions.
Consequently, approximately 1.5k resident tok/s from that path is an
\emph{approximate-target} result. It cannot be presented as native AR or
lossless full-target verification, even if an early France answer passed.

M128 is a row count, not inherently speculative: native C128 would also
produce M128, whereas native C32 normally produces M32. Padding C32 to 128
does not compute three useful future AR positions for free. The new native
down-consumer pilot therefore tests actual M32 without reintroducing DSpark.

\subsection{Strict TP8 MHC fusion checkpoint}

The September 11 trial preserved all target experts, original weights,
TP-synchronized draft/accept decisions, a 1M logical pool, and gamma three.
It used 32 real-code requests, 512 outputs with \texttt{ignore\_eos}, and
chunk 2304. Four measured resident windows per family gave:

\begin{table*}[tp]
  \centering
  \caption{Historical full-target DSpark only; not a current AR comparison.
  Acceptance and throughput are reported in their original trial scope.}
  \label{tab:dspark-history}
  \begin{tabular}{@{}lrrr@{}}
    \toprule
    Family & Resident tok/s & Mean accept length & Relative to control \\
    \midrule
    Control & 1073.16 & 2.687 & --- \\
    Fusion + FP32 mixing & 1127.48 & 2.732 & +5.06\% \\
    Fusion + FP16 mixing & 1142.26 & 2.698 & +6.44\% \\
    \bottomrule
  \end{tabular}
\end{table*}

FP32 mixing was selected historically. The FP16--FP32 mean gap of about
1.31\% was smaller than the observed FP16-family range of 2.5\%, their ranges
overlapped, and only four windows per family were available. FP16 also had a
larger component perturbation and the trial's single observed semantic
degeneration. These observations motivate the conservative choice, not a
statistical proof that either rounding variant is universally safer.

\begin{figure*}[tp]
  \centering
  \includegraphics[width=.90\textwidth]{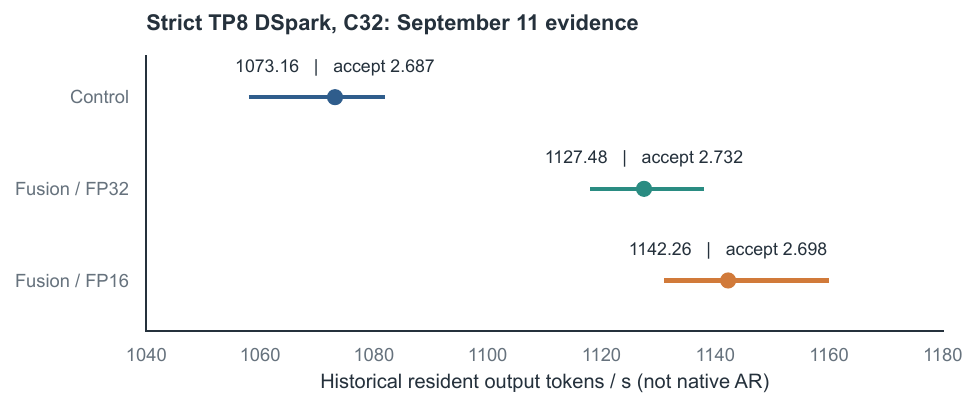}
  \caption{Historical strict DSpark family means with the ledger's rounded
  observed min/max ranges, not confidence intervals. Native AR, natural EOS,
  and the September 14 C32 pilot are deliberately not plotted on this axis.}
  \Description{Historical strict DSpark means are 1073.16 for control, 1127.48 for FP32 fusion, and 1142.26 for FP16 fusion; intervals are rounded observed ranges, not confidence intervals.}
\end{figure*}

\subsection{Acceptance, timing, and drift}

For concurrency $C$, average committed tokens per request $a$, and complete
speculative step time $T_{\mathrm{SD}}$, useful throughput is
$Ca/T_{\mathrm{SD}}$. It exceeds native AR only if
$T_{\mathrm{SD}}/T_{\mathrm{AR}}<a$. Larger verification batches expand expert
work and may not amortize draft cost. Acceptance alone is not a speed metric.

In the historical trial, chunked admission ramped the active batch through
several sizes before all requests were resident. Controls differed from
themselves on most complete output sequences. Such changes in batching and
numerical execution confound attribution from final hashes; they do not prove
that all other numerical or synchronization risks are absent. Most flagged
dot tails were forced continuation after EOS, distinct from genuine semantic
repetition. This motivated the natural-EOS protocol used in the new AR matrix.
The 1127.48 figure must not be divided by the new 1044.32 AR figure to claim a
matched speculative speedup: output policy, input mixture, chunking and date
differ. A fresh matched AR/strict-DSpark experiment remains outside this update.

\section{Measurement methodology}
\label{sec:method}

\subsection{Resource and metric controls}

Every GPU experiment begins with:

\begin{lstlisting}[language=bash]
amd-smi process --json
\end{lstlisting}

Launchers compare GPU owners with their recorded service PID, command, and
descendants before each test. The current campaign additionally checks Linux
boot ID and process start ticks, avoiding wall-clock-derived process identity.
They refuse unrelated GPU owners and terminate only the owned test process
tree. The native campaign verifies the original model path, TP size,
speculative algorithm being absent, actual graph row counts, and KV pool size.
Logged flags alone are not proof that a specialized kernel was selected.

Early experiments used NUMA node 1 because the host had a known node-0 memory fault, independent of MI250 or SGLang. The operator subsequently reported that fault repaired. The September 13--14 launcher uses NUMA interleaving across the host; it does not retain the old node-1-only policy. This report does not claim a new hardware-health diagnosis.

We sampled device memory with AMD-SMI at roughly five-second intervals.
The highest sampled use was \PeakMemoryMB{} of 65,520 tool-reported MB on a
GCD (\PeakMemoryPercent{}\%). These are sampled device observations, not exact
allocator peaks or occupied-KV counts. The setting
\path{mem_fraction_static=0.96} controls a budget, not constant 96\% utilization.
Logical KV capacity, allocated storage, graph and workspace memory, and
occupied context are different quantities.

\subsection{Timing definitions and aggregation}

Let $f_i$ and $e_i$ be the first and last streamed-token timestamps of request $i$, and $n_i(t)$ its cumulative output count. The common resident interval is
\begin{equation}
  t_s=\max_i f_i,\qquad t_e=\min_i e_i,\qquad
  R_D=\frac{\sum_i[n_i(t_e)-n_i(t_s)]}{t_e-t_s}.
\end{equation}
Only intervals with $t_e>t_s$ contribute. Multiple waves contribute token and duration sums within each formal round; the reported cell is the median of three round rates. This is streamed service wall time, not pure GPU timing. Whole-wave HTTP rate instead divides all completed output tokens by wave start-to-last-response time. Prefill rate divides all prompt tokens by wave start-to-last-first-token time, including admission and first-token overhead. The two-column P/D matrix does not describe disaggregated serving.

Candidate comparisons use the order

\begin{equation}
  A_1 \rightarrow B_1 \rightarrow B_2 \rightarrow A_2.
\end{equation}

A1 and A2 use independently started control processes; consecutive B1/B2 legs can share one candidate process. This is stated per experiment, not represented as four independent starts. Explicit warmup files are excluded. Measured outliers remain in the formal three-round matrix. The small down-consumer pilot uses two fixed waves per leg and mean arm medians for its ABBA comparison. No statistical confidence interval is inferred from these small samples. Earlier trimmed-mean or short-probe protocols retain their original labels.

The initial request can trigger compilation even after another shape was warm. The new corpus includes rows of 511 and 512 tokens; admission groups exposed M8190 and M8191 below the M8192 CK selector. Object timestamps and Ninja records associated four builds totaling about 23.1 s with one slow wave. This evidence explains that cold-shape event, not all HTTP variability. Runtime-M repair already removed one small-prefill specialization problem, but remaining exact-M tails are an open latency risk. Generated IDs, raw stream samples, finish reasons, input hashes, and cold/warm separation are retained outside temporary directories.

\subsection{Correctness hierarchy}

\begin{enumerate}
  \item Kernel level: elementwise bitwise equality, maximum absolute error, relative L2, and cosine similarity.
  \item Layer level: router top-k, stage 1, activation, stage 2, and rank-local sum.
  \item Model level: decode completion IDs independently and check semantic sentinels and logits on fixed inputs. France is a sanity check, not a quality benchmark.
  \item Attribution level: batch shapes, admission, prefixes, physical cache state, and numerical path must be controlled before comparing whole output hashes. Greedy divergence alone cannot identify the faulty operator.
  \item Remaining coverage: broad teacher-forced tests comparing cached decode
  with recomputation at compression, window, and indexer boundaries, plus
  filled-long-context quality evaluation. Existing local fixtures are not
  universal coverage.
\end{enumerate}

\section{Rejected or reverted directions}

The following historical observations apply to the tested shapes and configurations. They do not prove a general algorithm or ISA instruction is unusable, and were not all rerun during this report update. Component timing is never substituted for a service-level win.

\begin{table*}[tp]
  \caption{Important negative results and their implications.}\label{tab:negative}
  \begin{tabularx}{\textwidth}{@{}p{0.27\textwidth}p{0.22\textwidth}Y@{}}
  \toprule
  Direction & Result & Conclusion \\
  \midrule
  CKTile KSPLIT=2/4 & Approximately 25/22 tok/s or lower & Split-K reduction and CTA pressure do not suit BS1 small-M MoE. \\
  AsyncLL / SDMA Mori & Approximately 20--49 tok/s & The current low-latency transport does not shorten the critical path. \\
  Offline FP4-to-INT8 expansion & Gate 7.23$\rightarrow$10.45 ms & Doubled weight traffic costs more than removing online unpack. \\
  CK BF16 router instances & Approximately 86.7--98.6 $\mu$s & Slower than stable torch/rocBLAS; not integrated. \\
  Simple LDS BF16 MFMA router & Approximately 0.43--0.52 ms & Barrier and bank/layout costs dominate. \\
  32$\times$32 INT8 MFMA gate & Best approximately 7.87 ms & 165 VGPR plus 32 KiB LDS collapses occupancy. \\
  Full prefill CUDA Graph & Capture succeeds; replay faults & Prefill metadata addresses are not capture-stable. \\
  One-CTA/token HIP mHC prefill & Approximately 28\% slower & Scalar FMA cannot replace the current decomposition by launch reduction alone. \\
  RCCL in place of peer-read AR & TTFT 0.679$\rightarrow$0.711 s & Reduce collective boundaries instead of substituting RCCL. \\
  BF16 down partial & Approximately 0.8\% micro gain with broad value changes & Insufficient benefit; retain FP32 partials. \\
  \bottomrule
  \end{tabularx}
\end{table*}

\section{Serving, memory, and agent integration}

\subsection{Reproducible configuration, not a live deployment}

The earlier four-GCD deployment allocated a 560,896-token pool and exposed graph tiers 1/2/4. The latest tested configuration instead uses TP8/EP1, pool 1,048,576, memory fraction 0.96, seven decode tiers, admission 16 for prefill, and chunk 32,768. Benchmark services are shut down after testing. No current endpoint availability, LAN exposure, or continuously running GPU workload is claimed.

The documented launcher starts a loopback service on port 30021 only when explicitly invoked. A user should discover its served model ID before sending a chat request; the performance harness bypasses chat tokenization by supplying the frozen input IDs directly.

\begin{lstlisting}[language=bash]
curl http://127.0.0.1:30021/v1/models
curl http://127.0.0.1:30021/v1/chat/completions \
  -H 'Content-Type: application/json' \
  -d '{
    "model": "<served-model-id>",
    "messages": [{"role": "user", "content": "Hello"}],
    "max_tokens": 64
  }'
\end{lstlisting}

\subsection{Historical API and agent integration}

Earlier work implemented OpenAI-compatible model discovery, streaming chat, a separate optional system-prompt proxy, and tool-result round trips for DeepSeek Harness~\cite{dsh}. Those endpoints were not part of the new performance measurements, and the proxy's availability is not revalidated here. No private host address or persona is needed to reproduce the benchmark.

The earlier parser configuration was:

\noindent\begin{minipage}{\linewidth}
\begin{lstlisting}
--tool-call-parser deepseekv4
--reasoning-parser deepseek-v4
\end{lstlisting}
\end{minipage}

Streaming/non-streaming tool-call checks were interface smokes, not throughput or model-quality measurements. Reproduction should use the archived launcher and input manifest rather than infer configuration from an old endpoint example.

\section{Remaining risks and next steps}

\begin{enumerate}
  \item \textbf{Capacity is not occupancy.} Fill the allocated 1M pool with controlled long contexts and verify quality, admission, sustained throughput, and exact memory peaks. Current measurements do not provide this evidence.
  \item \textbf{Numerical coverage.} Preserve fixed Top-K membership and order, test real nonconstant scales, and distinguish reduction effects from batch scheduling effects. Unique-slot CK removes one source of nondeterminism, but different GEMM shapes can still perturb scores and membership. Tested repeatability does not imply universal batch invariance.
  \item \textbf{Cold-shape coverage.} Audit shape-specific compilation keys
  and pre-warm supported shapes. Report cold, restarted, and warm TTFT
  separately. Do not hide latency outliers inside resident-only speed claims.
  \item \textbf{Indexer work.} Large-prefill score/Top-K ownership is now accepted, but owner-only query production is not. A fixed-hipBLASLt producer matched a historical layer fixture yet failed live layer-2 validation: two query rows changed three Top-K members, producing 291 positional differences in sorted lists. Local Top-K and publication matched their score inputs; projection arithmetic was upstream of the discrepancy. This rejected candidate remains disabled in the current campaign.
  \item \textbf{Controlled comparisons.} Matched TP4 and TP8 matrices and a fresh strict-DSpark regression remain missing. This update ran neither; unavailable temporary artifacts are not reconstructed as invented measurements.
  \item \textbf{Deployment decision.} The C32 consumer remains opt-in after a small positive screen. Broader request mixtures and context ranges are required before changing a global default. The C1 isolation result is not another kernel optimization.
\end{enumerate}

\section{Conclusion}

The expanded study combines historical TP4 correctness recovery with a fresh
single-instance TP8 evaluation. The current native-AR matrix reaches
\DecodeOne{} output tok/s at C1, \DecodeThirtyTwo{} at C32, and
\DecodeSixtyFour{} at C64. For C16, the prefill rates are \PrefillEight{},
\PrefillSixteen{}, and \PrefillThirtyTwo{} input tok/s for approximately
8K, 16K, and 32K requests with a 1M allocated logical pool. Reduced replicated
work, mHC reuse, paired-GCD attention, and fixed-order expert writeback jointly
define this measured configuration. Historical ABBA percentages are not
summed or represented as a freshly rerun cumulative ablation.

The evidence emphasizes the mismatch between stored precision and execution formats, small expert matrices, redundant work, specialization reachability, and fixed synchronization boundaries. It does not provide a hardware-counter roofline proving a single universal bottleneck. Keeping native AR, strict speculation, approximate targets, HTTP latency, and resident throughput separate is as important as the kernel work itself. The final deliverable is an auditable configuration and evidence archive, not a claim of universal bitwise equivalence or a completed million-token quality benchmark.

\bibliography{references}

@misc{dsv4card,
  author = {{DeepSeek-AI}},
  title = {{DeepSeek-V4-Flash} Model Card},
  year = {2026},
  url = {https://huggingface.co/deepseek-ai/DeepSeek-V4-Flash}
}

@manual{amdisa,
  author = {{Advanced Micro Devices, Inc.}},
  organization = {Advanced Micro Devices, Inc.},
  title = {{AMD Instinct MI200 CDNA2} Instruction Set Architecture},
  year = {n.d.},
  note = {Accessed September 14, 2026},
  url = {https://www.amd.com/content/dam/amd/en/documents/instinct-tech-docs/instruction-set-architectures/instinct-mi200-cdna2-instruction-set-architecture.pdf}
}

@misc{dsv4report,
  author = {{DeepSeek-AI}},
  title = {{DeepSeek-V4}: Towards Highly Efficient Million-Token Context Intelligence},
  year = {2026},
  howpublished = {arXiv:2606.19348},
  url = {https://arxiv.org/abs/2606.19348}
}

@misc{dspark,
  author = {Cheng, Xin and others},
  title = {{DSpark}: Confidence-Scheduled Speculative Decoding with Semi-Autoregressive Generation},
  year = {2026},
  howpublished = {arXiv:2607.05147},
  url = {https://arxiv.org/abs/2607.05147}
}

@misc{sglang,
  author = {Zheng, Lianmin and others},
  title = {{SGLang}: Efficient Execution of Structured Language Model Programs},
  year = {2024},
  howpublished = {arXiv:2312.07104v2},
  url = {https://arxiv.org/abs/2312.07104}
}

@misc{vllm,
  author = {Kwon, Woosuk and others},
  title = {Efficient Memory Management for Large Language Model Serving with {PagedAttention}},
  year = {2023},
  howpublished = {arXiv:2309.06180},
  url = {https://arxiv.org/abs/2309.06180}
}

@misc{dsv3,
  author = {{DeepSeek-AI}},
  title = {{DeepSeek-V3} Technical Report},
  year = {2024},
  howpublished = {arXiv:2412.19437},
  url = {https://arxiv.org/abs/2412.19437}
}

@misc{megablocks,
  author = {Gale, Trevor and Narayanan, Deepak and Young, Cliff and Zaharia, Matei},
  title = {{MegaBlocks}: Efficient Sparse Training with Mixture-of-Experts},
  year = {2022},
  howpublished = {arXiv:2211.15841},
  url = {https://arxiv.org/abs/2211.15841}
}

@misc{ck,
  author = {{AMD ROCm}},
  title = {Composable Kernel},
  year = {n.d.},
  note = {Accessed September 14, 2026},
  url = {https://github.com/ROCm/composable_kernel}
}

@misc{aiter,
  author = {{AMD ROCm}},
  title = {{AIter}: {AI} Tensor Engine for {ROCm}},
  year = {n.d.},
  note = {Accessed September 14, 2026},
  url = {https://github.com/ROCm/aiter}
}

@misc{dsh,
  author = {{DeepSeek-AI}},
  title = {DeepSeek Harness},
  year = {n.d.},
  note = {Accessed September 14, 2026},
  url = {https://github.com/deepseek-ai/deepseek-harness}
}

\appendix

\section{Key checkpoints}

Table~\ref{tab:checkpoints} lists implementation and evidence revisions.
Dates and protocol boundaries are recorded in the corresponding experiment ledgers.

\begin{table*}[tp]
  \caption{Implementation and evidence checkpoints.}
  \label{tab:checkpoints}
  \begin{tabularx}{\textwidth}{@{}p{0.20\textwidth}Y@{}}
  \toprule
  Commit & Content \\
  \midrule
  \commit{505b3373794a} & Initial gfx90a/ROCm enablement \\
  \commit{caf80718cf} & CKTile FP4 expert W2 output-order repair \\
  \commit{9140c03605} & Correct TP-only 60 tok/s route \\
  \commit{b097d228c2} & Native router decode optimization \\
  \commit{ea519e2e0f} & Routed-MoE prefill MFMA \\
  \commit{6680714cf8} & Sparse-prefill wave reduction \\
  \commit{6547c5b063} & M2048 64-row expert block; long-prefill gain of 5.6\% \\
  \commit{85ab1a40c5} & DSV4 reasoning and tool parser wiring \\
  \commit{94d630a47c} & Full-endpoint OpenAI client compatibility \\
  \commit{b00a4e11cf} & Historical TP4 tree reproduced at 74.018 HTTP tok/s median on September 6 \\
  \commit{137170e367} & TP8 large-prefill migration with corrected scale interpretation \\
  \commit{9edb8ecdde} & Historical strict-DSpark FP32 MHC fusion profile (not the latest AR matrix) \\
  \commit{c224e9c1ce} & Fresh-start unified-V4 pool/metadata integration repair \\
  \commit{3d73a75923} & Formal seven-tier native P/D archive and C1 GEMV restoration \\
  \commit{fb2e39fca9} & Default-off native C32 down-consumer pilot and C1 isolation \\
  \bottomrule
  \end{tabularx}
\end{table*}

\section{Reproducible report build}

\begin{lstlisting}[language=bash]
cd reports/gfx90a-dsv4
python generate_figures.py
make check
make
make arxiv
\end{lstlisting}

The report follows the supplied PPoPP workshop template's
\texttt{acmart} class. Its options are \path{sigplan,10pt}; the
\texttt{review} option is deliberately omitted because the public arXiv
version must not contain line numbers. The class controls two-column layout,
margins, fonts, headings and captions;
no geometry or body-font override is applied. Bibliography formatting uses
\texttt{ACM-Reference-Format} and BibTeX. Figures embed Arial and retain only
left/bottom spines. Normal compilation needs neither Python nor a GPU.
The source bundle contains all TeX inputs, bibliography sources, the generated
\texttt{main.bbl}, and pre-generated PDF figures. Local compilation of the
extracted bundle is verified; arXiv's remote service is not tested here.
The 2024 workshop date is not used as a claim of publication: this is a
September 2026 technical report using that template, with no submission-page
limit implied.

\section{Evidence provenance and open gaps}

In the companion repository, \path{data/results.json} freezes three rounds per
formal cell, ABBA samples, hashes, and archive identifiers. The script
\path{prepare_report_data.py} refreshes that snapshot from repository evidence;
normal report builds do not access parent directories. The two repository
evidence directories are:
\begin{itemize}
  \item \path{.agents/experiments/dsv4_tp8_revalidation_20260913/}
  \item \path{.agents/experiments/dsv4_tp8_ar_down_consumer_20260914/}
\end{itemize}
The formal archive contains 278 verified files; the pilot contains 92. Complete SHA256 values are retained with the report data. These checks protect recorded artifacts, not unrecorded measurements or full checkpoint weights.

Historical TP4 records and the September 11 DSpark table are ledger evidence with explicitly different protocols. Older temporary full-matrix files were unavailable during this update, so no missing TP4 concurrency cells are filled from memory. Public source excerpts are fixed at repository revision \commit{54b93c45c2}; a source-path list may be longer than the text retained after prompt trimming. Neither source-review prose nor a simple France question measures general code correctness. Dataset-specific performance should not be mistaken for a model-quality benchmark or a matched comparison against another hardware platform.

\end{document}